# Response to the authors of "On the (un)effectiveness of Proton Boron Capture in Proton Therapy"

GAP Cirrone[1], G Cuttone[1], L Manti[2], D.Margarone[3], G Petringa[1], L.Giuffrida[3], A.Minopoli[2], A.Picciotto[4], G.Russo[5], F.Cammarata[5], P.Pisciotta[1], F.M.Perozziello[2], F.Romano[1,6], V.Marchese[1], G.Milluzzo[1], V.Scuder[1,3]i, G.Cuttone[1] and G. Korn[3]

[1]Istituto Nazionale di Fisica Nucleare- Laboratori Nazionali del Sud,
via S. Sofia, 62, Catania, Italy.

[2]Physics Department, University of Naples Federico II, Naples, Italy.

[3]Institute of Physics ASCR, v.v.i. (FZU), ELI-Beamlines Project,
Na Slovance 2, Prague, 18221, Czech Republic.

[4]Fondazione Bruno Kessler, Micro-Nano Facility, Via Sommarive 18,
38123 Povo-Trento, Italy.

[5]Institute of Molecular Bioimaging and Physiology - National Research Council -
(IBFM-CNR), Cefalù, (PA), Italy.

[6]National Physical Laboratory, Acoustic and Ionizing Radiation Division,
Teddington, TW11 0LW, Middlesex, United Kingdom.

**Abstract**

This manuscript provides a response to a recent report by Mazzone et al. available online on arXiv that, in turn, tentatively aims at demonstrating the inefficacy of proton boron capture in hadrotherapy. We clarify that Mazzone et al. do not add any scientific or technical insights to the points extensively discussed in the original manuscript by Cirrone et al., and/or in the series of iterations had with the Referee, which ultimately lead to the publication of our original and pioneering experimental work. Here we summarize some of the key points of the long scientific debate we had during the review process of paper by Cirrone et al., which are very similar to the considerations presented by Mazzone et al.. In conclusion, no quantitative explanation of our robust experimental achievements presented in Cirrone et al. is provided in Mazzone et al.

**Introduction**

A work recently published by our group [1] experimentally demonstrates for the first time the potential role of the p+11B -> 3a (for brevity, p-B) reaction in the biological enhancement of proton therapy effectiveness. The work reports robust experimental data in terms of clonogenic curves and chromosomal aberrations and unambiguously shows the enhancement when cells were exposed to a clinical proton beam subject to treatment with sodium boroncaptate (BSH). Moreover, the larger occurrence of complex-type of chromosomal exchanges highlights the presence of an LET larger than that of protons alone, possibly related to alpha-particles generated by the pB fusion reaction (see for instance Sumption et al., 2015)

At the same time, we have openly and clearly reported in [1] that analytical calculations, performed on the basis of the well known total production cross section curves, are not able to explain the effect in a macroscopic way, i.e. solely in terms of a trivial increase in the total dose presumably released in the cells by the generated alpha-particles.

In this document, we report all the steps of the scientific discussion carried out during the review process of Scientific Reports based on the points raised by the Referees. This includes several calculations and considerations exchanged with the Referees between April 2017 and December 2017, i.e. well before our publication [1].

The conclusion of such a long scientific debate with the Referees was that, however desirable, we were not able to provide a simple analytical computation to explain quantitatively our experimental results, for instance by correlating the biological effect with the total number of α-particles that is expected in our experimental conditions.

A work recently published on arXiv by Mazzone et al. [2], just one month after our publication, reports trivial calculations on p-B generated alpha-particles, incidentally very similar to those privately exchanged between us and the reviewers. The authors of [2] have no choice but concurring on what we already admitted with the reviewers of [1], reaching the same conclusion as ours on the well known involved nuclear processes, while adding nothing on the debate we started. An alternative interpretation of our experimental data, with sound physical bases and a thorough understanding of the complexity of the underlying radiobiology, is totally missing in [2]. This would have been much more useful to us and the scientific community interested in the potential improvement of protontherapy efficacy, experimentally demonstrated for the first time by our group as reported in [1].

The only apparent "novelty", although misleading, added by the authors of [2] is the introduction of an argument based on the p-16O reaction that, as well known, is an important channel related to the production of alpha particles in water. They report (Figure 2 in Mazzone et al.) the probability per unit length of proton interaction with water elements, including BSH at the density used in our experiment [1]. It is trivial that alpha-particles generated in the p+16O reaction dominate with respect to those generated via p+11B fusion, and this is obviously due to the high and natural concentration of Oxygen in water. Furthermore, this is also clearly misleading since BSH does not contain 16O. Figure 2 of Mazzone et al. is hence intended to deceive, leading the reader to think that the presence of oxygen in water (or in the human body) is responsible for alpha particle generation and dose increase. Alpha particles generated by protons interaction with Oxygen (or any other element naturally contained in the body), as well known by any expert in the field, are present and obviously already studied in the development of biological treatment planning for protontherapy. This has not related at all with the presence of BSH (or boron) in the cells used in our experimental runs and, in any case, it is an effect always present (with and without BSH or B).

Data reported in this work aim at clarifying that the discussed radiobiological phenomena in our paper [1] will have to be better explained from an analytical/numerical point of view and, at the same time, that their interpretation cannot be merely carried out with a classical, non sub-micrometric approach.

**p+11B scientific literature related to Medical Physics applications**

In 2014, a paper by Y. Do-Kun et al [3] reported on Monte Carlo simulations supporting the use of p + 11B → 3α nuclear fusion reaction to enhance proton biological effectiveness exclusively in the tumour region through the generation of short-range high-LET alpha particles. They also proposes to develop a tumor-treatment monitoring technique using prompt gamma rays emitted from 11B atoms interaction with incident protons [4-9].

In the subsequent period we have published a number of theoretical papers [7-9] clarifying some points on the dose enhancement presented in [3] (in fact by proposing a few corrections, especially based on the fact that the authors of [3] had used unrealistic B concentrations in the body) and on the origin of the prompt gamma rays (10B and not 11B).

On January 18th, 2018 we published the first pioneering experimental data [1] demonstrating in vitro the predicted effect and suggesting the role of the p+11B reaction in the observed enhancement of proton-induced clonogenic cell death and DNA damage. In particular, we concluded our paper with the sentence: "The radiobiological data reported in this work suggest that the p + 11B → 3α reaction as being responsible for the observed increase in the biological effectiveness of a clinical proton beam. However desirable, we cannot currently provide a simple analytical computation able to explain our results, for instance by correlating the biological effect with the total number of α-particles that can be expected to be generated under our experimental conditions." The discussion on the macroscopic demonstration of the effect is, hence, left open to further investigation.

On February 26th, 2018 the paper by Mazzone et al. [2] appeared on arXiv. The manuscript reports obvious physical considerations already raised in the discussions we previously had with the Referees of Scientific Report during the review process of [1]. Furthermore, remarks presented in [2] are very similar to some of the discussions carried out with the Referees both in terms of content and form.

Based on this premise and for the sake of clarity, we here report the outstanding aspects of the debate we had with the Reviewers of Scientific Reports, where almost all calculations reported in the paper by Mazzone et al. where already anticipated. We also underline that the analytical considerations we had carried out and discussed were removed from the original paper, in accordance with the Referees' request.

The discussions reported in the following paragraph have an important scientific value as they emphasise all the issues still to be solved and understood in such recently emerging field.

**Summary of the Correspondence with the Scientific Reports' Referees on the quantification of the generated alpha particles**

We here report the summary of four different iterations we had with the Scientific Reports' Referees who had sent us their comments on April 3rd, July 20th, November 8th and December 15th 2017. This summary refers exclusively to the discussion raised by one of the Referee on the the alpha particle estimations and on its potential inconsistency with the corresponding radiobiological estimates from our experimental results.

Since the first interaction and for all the subsequent reviews, one of the Referees criticised our work for the lack of a dosimetric evaluations i.e. for the estimation of the dose released by the alpha particles produced in the considered reaction. In fact, the omission of these considerations, in the first manuscript version, was intentional: the work was conceived as a proof-of-principle, with the only goal to demonstrate/verify experimentally the biological effect.

The statement that the dose released is one-to-one connected with the corresponding biological effect, is wrong as easily understandable by an expert in the field. The mere macroscopic concept of absorbed dose, in fact, does not allow to explain alone the biological effects of radiations that, on the opposite, must be analysed by the use of micro and nano dosimetric approaches. Nevertheless, at that time, we decided to show the Monte Carlo and analytical calculations we had already performed, in the attempt to explain with a trivial approach the observed effect.

Using the Geant4 Monte Carlo Hadrontherapy [10-14] application we have simulated the beamline in detail and then evaluated the the proton and neutron spectra at the cell position. Using the well known total cross section of the 11B(p,a)2a reaction [15-16] we then evaluated the number N of total alphas produced with the formula:

$$N = \frac{\varrho \cdot \Delta x \cdot N_A}{A} \sum_{E=1}^{n} N_E^p \cdot \sigma(E) \quad (1)$$

where the experimental total reaction cross section σ(E) is those reported in Figure 1 [15-16], the Boron density is assumed to be 8E-5 gr/cm3 (corresponding to the one used during the experiment) and the layer cells thickness is assumed to be 100 um.

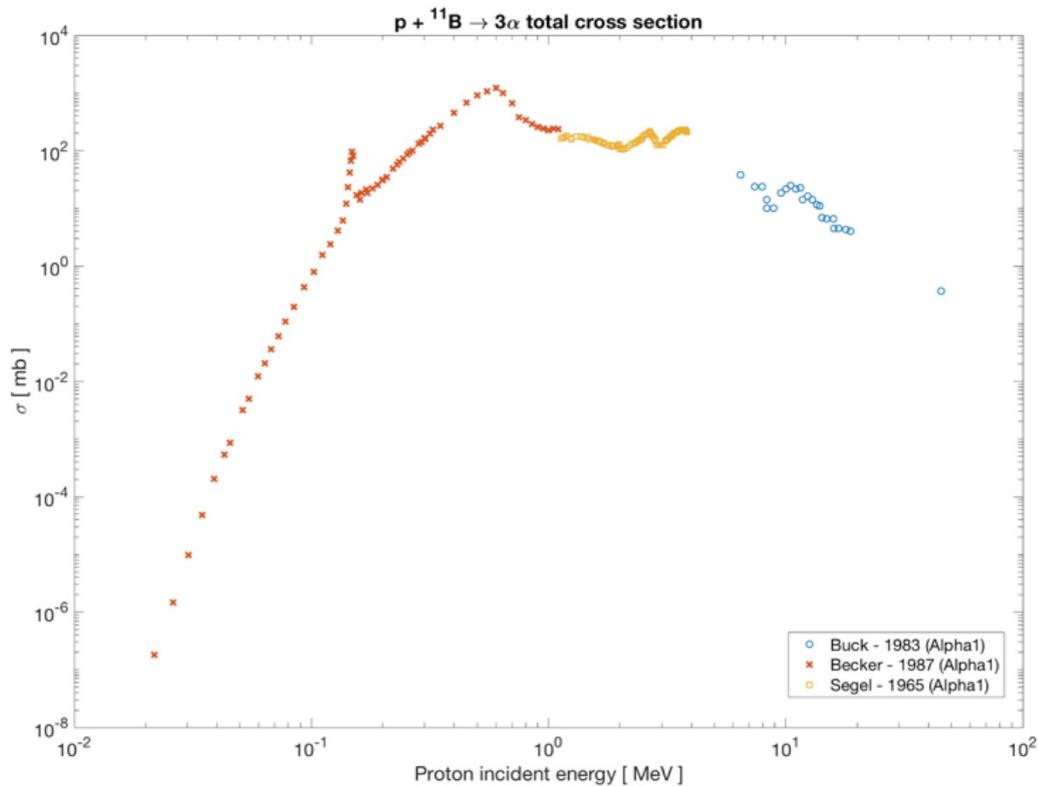

Figure 1: experimental total cross section of the 11B(p,a)2a reaction.

The considered reaction shows a maximum cross section of 1.2 b at the proton energy of 675 keV, as reported in Figure 1. The number of alpha particles as a function of the proton depth in water is reported in Figure 2 along with the experimental Spread Out Bragg Peak.

Moreover, Monte Carlo simulations allowed us to estimate the increment of physical dose due to alpha particle production. The resulting alpha particle dose is of the order of $10^{-7}$ Gy in the middle of the SOBP, thus apparently negligible with respect to the primary proton dose. For the sake of clarity Figure 2 shows the absolute dose spread out Bragg peak when 8.47E11 primary proton histories are simulated. In the same figure the correspondent absolute dose released by the generated alpha particles is shown.

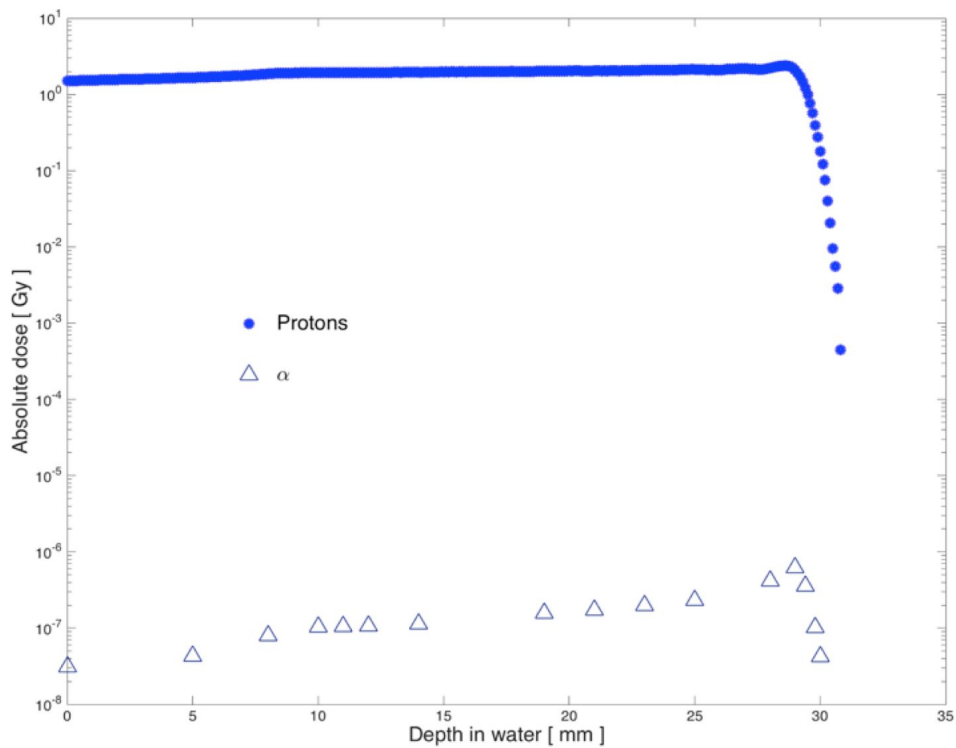

**Figure 2:** Simulated absolute dose released by the proton beam and by only alpha particles (log-scale)

As the reader can easily argue, our first estimation of the total number of alpha particles produced was already performed in April 2017. It should be noted that Figure 2 is very similar to Figure 6 of the paper by Mazzone et al [2].

During our several iterations with the Reviewers, lasting for about nine months (April-December 2017), we have also discussed and evaluated other potential source of alpha particles. Among these, we have discussed the contribution of neutron interaction with 10B based on the very well known 10B(n,7Li)a channel, the presence of the 10B(p, a)7Be channel and the possibility of a locally increased density of BSH in the cells with respect to the medium reported in literature [17].

We concluded our various responses admitting our obvious difficulty in finding an analytical explanation of our robust experimental results. On the contrary, considering the robustness of our experimental data and given the vast uncertainties in terms of boron localization within the cells (similarly to the 70-year old history of BNCT), we expressed our conviction that this point will be elucidated in future theoretical papers by the international community of experts in the field. Unfortunately the Referee was against this position and concluded in December 2017 that the Authors were not able, at this stage, to demonstrate the effect and that this should have been clearly emphasised in the conclusion. We accepted such a Referee's final remark and wrote a sentence explicitly and clearly admitting that the role of p-11B reaction in the work is probable but remains an hypothesis that must be still investigated.

The final sentence reported in our original paper is [1]: "The radiobiological data reported in this work suggest that the $p\ +\ 11B\ \rightarrow\ 3\alpha$ reaction as being responsible for the observed increase in the biological effectiveness of a clinical proton beam. However desirable, we cannot currently provide a simple analytical computation able to explain our results, for instance by correlating the biological effect with the total number of $\alpha$-particles that can be expected to be generated under our experimental conditions. A possible approach could be trying to calculate the increase in the overall dose and/or LET due to such particles. The current knowledge of biological radiation action has nonetheless established that, the biological effects of low-energy high-LET radiations cannot be interpreted solely on the basis of macroscopic concepts like the absorbed dose or the average LET distributions. This is due to the intrinsically inhomogeneous nature of energy deposition events along radiation tracks, which becomes more significant with their increasing ionization density. Therefore, micro- and nano-dosimetric approaches must be taken into account to analyse the effects arising at cellular level. In addition, the role of extra-targeted phenomena, such as the bystander effect whereby cells that have not been traversed directly by radiation tracks may express cytogenetic damage, is still largely undetermined in such scenario, thereby contributing to the overall uncertainty between the physical dose distribution at the micro or nano-dosimetric level and at the cellular one".

**Conclusions**

The aim of this paper is to openly show all the analytical and Monte Carlo studies performed during the preparation of our original paper [1] and during its review process, hence well before the publication of the paper by Mazzone et al [2]. These calculations were explicitly requested by one of the two Referees in the attempt of explaining the observed enhancement of a clinically used proton beam by means of classical arguments connected to the concept of the released dose. We decided to present these results fully and reporting the original Referees' questions and authors' replies in the hope of a better clarification on the fact that trivial and classical calculations performed with a nuclear approach are not sufficient to explain the experimentally observed effect.

The work proposed by Mazzone et al., besides the fact that does not provide any new scientific insights with respect to what already stated in our original paper, is limited to a number of trivial considerations, sometimes misleading, that attempt to discredit our pioneering experimental achievements, rather than contributing to its interpretation in an original and/or constructive way. Moreover, the statement on the role of p-16O reaction and of the possible radiosensibilisation of BSH via a not better specified "purely biochemical effect" (quoting Mazzone et al, page 8 of the Discussion section") clearly shows the lack of scientific arguments in their tentative to explain our results, thus leading to a rather superficial vision of the authors of [2].

Mazzone et al. compulsively put the emphasis only on the number of total produced alpha particles and on the macroscopic concept of absorbed dose, completely forgetting to mention that, for the interpretation of phenomena at cellular level, this approach is absolutely insufficient. Dose is by definition a macroscopically non-stochastic quantity. However, at the nanoscale level, i.e. that of relevance for DNA damage, hence for cellular effects, the concept of absorbed dose falls short of explaining exhaustively a large number of phenomena as the intrinsically discontinuous pattern of energy deposition by charged particles becomes important, leading to a inherently inhomogeneous biologically effective "dose" . With regard to this, they completely miss to mention and discuss the role that microscopic quantities such as the LET, play at this level demonstrating either a complete lack of basic knowledge in the field or a visible *mala fides*. Or both.

Knowing that the physical considerations on the number of produced alphas reported by Mazzone et al, besides trivial and obvious, are correct and shared by us, we hope that the next work on this important issue will not be written for the sole purpose to discredit another paper but with the desire (and capacity) to better understand and find explanations of these new radiobiological phenomena.